\documentclass[preprint,nofootinbib,12pt,floatfix,superscriptaddress,aps,prd]{revtex4}
\usepackage{amsmath}
\usepackage{amssymb}
\usepackage{amsbsy}
\usepackage{amsfonts}
\usepackage{amsopn}
\usepackage{amstext}
\usepackage{graphicx}
\usepackage{amssymb}
\usepackage{amsfonts}
\usepackage{amsmath}
\usepackage{graphicx}
\usepackage{color}
\usepackage{slashed}
\usepackage{esint}
\usepackage[dvips]{epsfig}
\usepackage[dvips]{graphicx}
\usepackage{float}
\usepackage{units}
\usepackage{textcomp}
\newcommand{\beq}{\begin{equation}}
\newcommand{\eeq}{\end{equation}}
\newcommand{\bea}{\begin{eqnarray}}
\newcommand{\eea}{\end{eqnarray}}

\begin{document}

\title[Four-dimensional regular black strings in bilocal gravity]{Four-dimensional regular black strings in bilocal gravity}


\author{C. R. Muniz\footnote{E-mail:celio.muniz@uece.br}}\affiliation{Faculdade de Educa\c{c}\~{a}o, Ci\^{e}ncias e Letras de Iguatu, Universidade Estadual do Cear\'a, Campus Multi-Institucional Humberto Teixeira, S/N, 63502-253, Iguatu-CE, Brazil.}
\author{H. R. Christiansen\footnote{E-mail:hugo.christiansen@ifce.ce.br}} \affiliation{Departamento de F\'{i}sica,  Instituto Federal de Educação, Ciência e Tecnologia do Ceará - IFCE , Av. Dr. Guarany, 317, Derby Clube, 62040-030,  Sobral-CE, Brazil.}
\author{M. S. Cunha\footnote{E-mail:marcony.cunha@uece.br}}\affiliation{Centro de Ci\^encias e Tecnologia, Universidade Estadual do Cear\'a, Av. Dr. Silas Munguba, 1700, 60714-903, Fortaleza-CE, Brazil.}
\author{J. Furtado\footnote{E-mail:job.furtado@ufca.edu.br}}\affiliation{Centro de Ci\^{e}ncias e Tecnologia, Universidade Federal do Cariri, 63048-080, Juazeiro do Norte-CE, Brazil.}


\begin{abstract}
In this paper, we obtain a static black string solution for a bilocal gravitational source in 3+1 dimensions. The solution is regular at the origin and tends asymptotically to the ordinary static uncharged black string solution of general relativity. It allows an event horizon and an internal horizon depending on the value of the mass density. A mass remnant associated with a vanishing Hawking temperature is also found. In order to stabilize the solution, a perfect cosmological fluid with state parameter $\omega>-1$ should be present throughout space. However, energy conditions do not exclude an exotic substance nearby the black string. Finally, we find the stationary counterpart of the solution and analyze the behavior of the horizons according to the mass and angular momentum of the spinning object.
\end{abstract}




\maketitle

\section{Introduction}
Space-time is presumably of a granular discrete nature at ultra-high energies. For this and other reasons quantum gravity has always been particularly difficult to connect with general relativity. Regularization procedures did not work in this realm and nonlocality was for a long time the only way to obtain a regular theory for useful calculations \cite{nonlocal, deserVN74}.
{In fact, it is the effective action with the effect of quantum fluctuations which is nonlocal, and provided it comes from a local action causality is guaranteed. \footnote{{The necessity of causality to be a feature of a fundamental theory can be argued at ultra high energies; however, its discussion is beyond the scope of the present report.}}}

In this paper we look for new regular black string solutions in a 3+1 dimensional space-time. For this goal one should first better understand black holes in 2+1 dimensions. It is known that there is a close connection between 2+1 gravity and pure Chern-Simons gauge theory \cite{witten88} but nothing like this has been found in 3+1 dimensions. On the other hand, the 2+1 Einstein-Hilbert theory at the classical level is trivial and the vacuum has no local degrees of freedom \cite{leutwyler}. This is because in 2+1 the Weyl tensor is zero and the Ricci tensor vanishes due to the Einstein field equations. As a result, the full Riemann tensor is trivial and no black hole solution with event horizons exists with $\Lambda\geq0$ \cite{ida2000}. Thus, the only possibility  of a three-dimensional singularity is to admit  $\Lambda<0$ giving rise to a solution known as a BTZ black hole \cite{BTZ}. In this case the BH has similar properties to the (3+1)-dimensional Schwarzschild and Kerr black holes, including their thermodynamic properties.   Most significantly, the BTZ BH radiates at a Hawking temperature and has an entropy equal to a quarter of its event horizon area. Recently, in an extension considering nonlocal gravity we have found two horizons and a mass remnant in the 2+1 static uncharged case \cite{ourpaper}.

Black strings are higher dimensional solutions of the Einstein equations in D-dimensional spacetime \cite{duff1988}.
These vacuum objects are symmetric generalizations of black holes under translations where the event horizon is topologically equivalent to $S_2 \times R$ (or $S_2 \times S_1$ in the case of black rings) and spacetime is asymptotically $M_{D-1} \times S_1$ for a zero cosmological constant or $AdS_{D-1} \times S_1$ for a negative one \cite{emparanPRL,ijmpa2011}.
A black string is a special case of a black $p$-brane solution (with $p = 1$), where $p$ is the number of additional spatial dimensions to ordinary space. A $p$-brane thus spawns a $(p+1)$-dimensional world-volume in spacetime.
In string theory, a black string describes a D1-brane surrounded by a horizon. It is well known that as an open string propagates through spacetime, its endpoints are required to lie on a {D-brane} on which it satisfies Dirichlet boundary condition, and the dynamics on the D-brane worldvolume is described by a gauge field theory \cite{polchinski95}. In models in which our universe is a world-brane in a higher dimensional space, the gravitational collapse of  matter would produce a
black hole on the brane. In asymptotically flat spaces, black string solutions are however unstable to long wavelength perturbations since the localized black hole is entropically preferred to a  long segment of string. The string’s horizon therefore has a tendency to form a line of black holes.  However, in AdS the space acts like a confining box which prevents fluctuations of long wavelengths from developing.  

Regarding low dimensional gravity, in Ref. \cite{jhep2000 b} it has been found that large black holes on a {2-brane} with a negative cosmological constant agree exactly with the 2+1 dimensional BTZ solution. Such black-holes are not localized on the brane but extend across the fourth dimension giving rise  to a BTZ black string. When the size of the extra dimension is finite, large black holes on the brane correspond to black strings in the bulk. This is what one expects from a standard Kaluza-Klein compactification. For smaller masses, we have two different black hole solutions in which the horizon does not reach the second brane. In both cases, the solution looks like the BTZ black hole with correction terms that fall off exponentially in the proper distance. For black holes on asymptotically flat two-branes, it was shown that the mass measured asymptotically on the brane agreed with a four dimensional thermodynamic mass obtained by integrating the relation $dM = TdS$ in the bulk \cite{jhep2000 a}. After that, Emparan, Horowitz and Myers have shown that the same is true for black holes on asymptotically AdS3 branes.

The paper is organized as follows: In section II we obtain the bilocal solution of a four-dimensional black string and discuss its properties, such as horizons, Hawking temperature, mass remnant, energy conditions and a stationary solution for a spinning black string. In section III we present our conclusions.

\section{Bilocal black string solution and its properties}

Quantum gravity based on a local field theory is nonrenormalizable at any loop order. As an alternative, nonlocal gravity models were defined by modifying the Ricci scalar term in the Einstein-Hilbert action \cite{nonlocalgrav}. Moffat and co-workers applied a model developed for electroweak interactions \cite{moffat qft} to handle quantum gravity in 3+1 dimensions
\cite{moffat gravity,moffat etal2011,NLAction}. The basic point is to shift the coupling constant by $\sqrt{G}= \mathcal{F}\sqrt{G_N}$, where $\mathcal{F}$ is an entire function \cite{moffat gravity}. This is in order to regularize the theory at ultra-high energies and satisfy unitarity to all orders.  In this approach, it is not the Einstein-Hilbert action but the matter action $S_M$ that incorporates the nonlocal coupling to gravity.
The generalized energy-momentum tensor is given by
\beq S_{\mu\nu} = \mathcal{F}^2(\Box/2\Lambda_G^2)T_{\mu\nu} \label{Snl}\eeq
where $\Box$ is the generally covariant d'Alambertian, $T_{\mu\nu}$ is the stress-energy-momentum tensor in the standard local coupling, and $\Lambda_G$ is a fundamental gravitational energy scale. The generalized Einstein field equations are
$ G_{\mu\nu} =8\pi G_N  S_{\mu\nu}$ where  $G_{\mu\nu}=R_{\mu\nu}-\frac12g_{\mu\nu}R$, as usual.

Therefore, the nonlocal differential operator produces a smeared energy-momentum tensor and the Einstein field equations describe gravity coupled to a generalized matter source term which includes full nonlocality.
Since $\mathcal{F}$ is invertible, the generalized Einstein field equations can be written as
\beq
\mathcal{F}^{-2}(\Box/2\Lambda_G^2)G_{\mu\nu} =8\pi G_N  T_{\mu\nu} \label{einstein2}
\eeq
In this way, Eq. \eqref{einstein2} reorders Eq. \eqref{Snl} having the same meaning but acting directly on the Einstein tensor. 
{Thus, the procedure we will adopt is to modify the energy-momentum tensor via the non-local operator given in Eq. (\ref{Snl})}.

Assuming that $z$, along the third spatial dimension where the black string is situated, can be factorized, the nonlocal 3+1 dimensional gravitational action is
\begin{equation}
	{S_{grav}=\frac{1}{2\kappa}\int d^4 x\sqrt{g}( \mathcal{R}(x)-\Lambda)}
\end{equation}
where in place of the Ricci scalar $R(x)$ we have
\begin{equation}
	\mathcal{R}(x)=\int d^3y  \sqrt{g}\mathcal{F}^{-2}(x-y)R(y).
\end{equation}
Here $\mathcal{F}^{-2}(x-y)=\mathcal{F}^{-2}(\Box_x)\delta^3(x-y)$ is a bilocal operator and $\Box_x$ is the generally covariant d'Alambertian.
The modified Einstein's equations become
\begin{equation}\label{EME}
	G_{\mu\nu}+g_{\mu\nu}\Lambda=\kappa\mathcal{T}_{\mu\nu},
\end{equation}
where $\mathcal{T}_{\mu\nu}=\mathcal{F}^{2}(\Box_x)T_{\mu\nu}$ and $\kappa=8\pi G_N/c^4$ is the Einstein gravitational constant.

\subsection{Static solution}

We will consider a static source in which $T_{0}^{0}=-\rho({\bf x})=-\mu\delta^2({\bf x})$, where $\mu=\int \rho({\bf x})r d\phi dr$ is the linear density of mass of the black string.
To see how it is modified by the vertex function, we compute
\begin{equation}
	\mathcal{F}^{2}(\Box_x)\delta^2({\bf x)}=(2\pi)^{-2}\int d^2p\, \mathcal{F}^{2}(-\nabla^2)e^{i{\bf p\cdot x}}=(2\pi)^{-2}\int d^2p\, \mathcal{F}^{2}(p^2)e^{i{\bf p\cdot x}}
\end{equation}
in a free-falling (Minkowisky's) reference frame. In theories of this kind we try to find a suitable form of $\mathcal{F}^{-2}(\Box_x)$ so that the
Euclidean momentum space function $\mathcal{F}^{2}(p^2)$ plays the role of the cutoff function in the quantum gravity perturbation
theory expanded against a fixed background spacetime at all orders. The bilocal operator is an entire function of order higher than or equal to 1/2. We will consider the simplest case 1/2, so that $\mathcal{F}(\Box_x)=\sqrt{4\pi}e^{\frac{1}{2}\ell\ \Box_x^{1/2}}$, where $\ell^2=-\Lambda^{-1}$ is a characteristic length (the constant $\sqrt{4\pi}$ was introduced just by convenience). Then, the modified energy density is \cite{Nicolini:2012eu,2014jwq}
\begin{equation}\label{ModifiedDensity}
	\tilde{\rho}({\bf x})=\mu\mathcal{F}^{2}(\Box_x)\delta^2({\bf x})=\frac{\mu}{4\pi(2\pi)^2}\int d^2p\,e^{-\ell p}e^{i{\bf p\cdot x}}.
\end{equation}


For cylindrical symmetry around the $z$ axis, in polar coordinates we get
\begin{equation}\label{Rho}
\tilde{\rho}(r)=\mu\int_0^{\infty }dp\,p\, e^{-\ell p}J_0(p r) =\frac{ \mu\ell}{\left(r^2+\ell^2\right)^{3/2}}.
\end{equation}
Since it represents the modified energy density now smoothed by the bilocal operator. It is worth to notice that the choice of the entire function is determinant.  The order of $\mathcal{F}$ would change the form of the energy density profile, as it happens with the $2+1$ regular black hole solutions discussed in \cite{ourpaper}. We now use Eq.(\ref{EME})
for the $00$-component {in which $\mathcal{T}^{0}_{0}=-\tilde{\rho}(r)$}, with the metric in the form
\begin{equation}\label{BlackStringMetric}
ds^2=-f(r)dt^2-\frac{dr^2}{f(r)}+ r^2d\phi^2+ \frac{(r^2+\ell^2)}{\ell^2} dz^2.
\end{equation}
The coefficient of $dz^2$ was chosen in order to stay flat at the origin and go like $r^2/\ell^2$
in the $r\to \infty$ limit.
For $f(r)$, we obtain
\begin{eqnarray}\label{PotMetr}
f(r)=&&\frac{\left(\ell^2+r^2\right)}{(\ell^2+2r^2)^{3/2}}\times
\Bigg\{\beta+ 2 \left[-\frac{\Lambda}{6}\left(\ell^2+2 r^2\right)^{3/2}+\kappa  \mu\ell \sqrt{\frac{\ell^2+2 r^2}{\ell^2+r^2}}\right.\nonumber\\
&&\left. - \frac{\kappa \mu \ell}{\sqrt{2}} \log \left(2\sqrt{2} \sqrt{\ell^2+r^2} \sqrt{\ell^2+2 r^2}+ 3\ell^2+4 r^2\right)\right]\Bigg\},
\end{eqnarray}
where $\beta$ is an integration constant which is fixed in order to make regular the solution at the origin and  anti-de Sitter for $r\ll \ell$. The result is
\begin{equation}
\beta= \kappa\mu\ell\left[\sqrt{2} \log {\left(2 \sqrt{2} \ell^2+3\ell^2\right)}-2\right],
\end{equation}
and thus $f(r)\approx 1+\frac{r^2}{\ell^2} \left\{1+14\kappa  \mu-6\sqrt{2}\kappa\mu\log{[\ell^2(3+2\sqrt{2})]}\right\}$ near the origin, where we considered $\ell^{-2}=-\Lambda/3$. The modified Ricci scalar at the origin is given by
\begin{equation}
R(0)=\frac{6 \left\{6 \sqrt{2} \kappa \mu \log \left[\left(3+2 \sqrt{2}\right) \ell^2\right]-14 \kappa \mu-1\right\}}{\ell^2},
\end{equation}
which confirms that the solution is singularity-free in a nonlocal theory in which there exists a fundamental length.

In the asymptotic limit {($r\gg \ell$)} we obtain
\begin{equation}\label{AsymptoticMetric}
f(r)\approx\frac{\ell \left[\sqrt{2} \kappa \mu \log \left(2 \sqrt{2} \ell^2+3 \ell^2\right)-2 \kappa \mu\right]}{2 \sqrt{2} r}+\frac{r^2}{\ell^2},
\end{equation}
which matches the static uncharged black string of general relativity for
\begin{equation}
\frac{2 \mu-\left[\sqrt{2}  \mu \log \left(2 \sqrt{2} \ell^2+3 \ell^2\right) \right]}{2 \sqrt{2}}\to 4\mu
\end{equation} \cite{Lemos:1995cm}.

\begin{figure}
\centering
 	\includegraphics[width=0.6\textwidth]{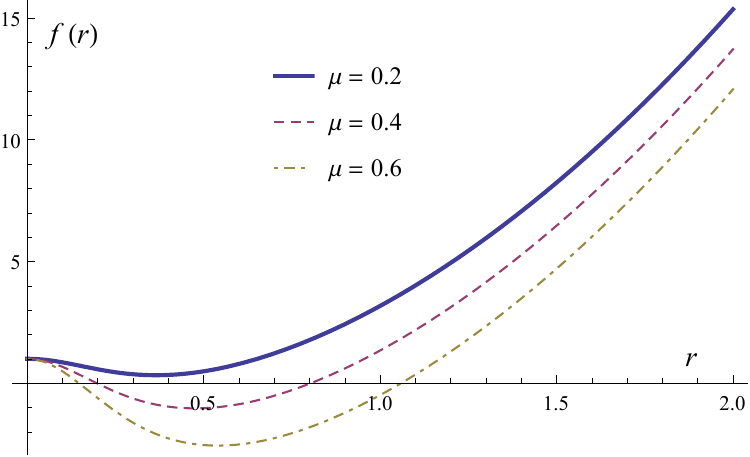}
 	\caption{Metric potential of the bilocal static black string as a function of the radial coordinate for some values of $\mu$, with $\ell=0.5$ and $\kappa=8\pi$.}
 	\label{effedeerre}
 \end{figure}

In Fig. \ref{effedeerre}, we depict the metric potential $f(r)$ for some values of $\mu$. Notice the existence of up to two horizons; the external (internal) one gets more distant (closer) from the origin the higher the value of the mass density is. Note also the regular behavior of the solution at the origin.

\subsection{Evaporation and remnant}
\begin{figure}
\centering
 	\includegraphics[width=0.6\textwidth]{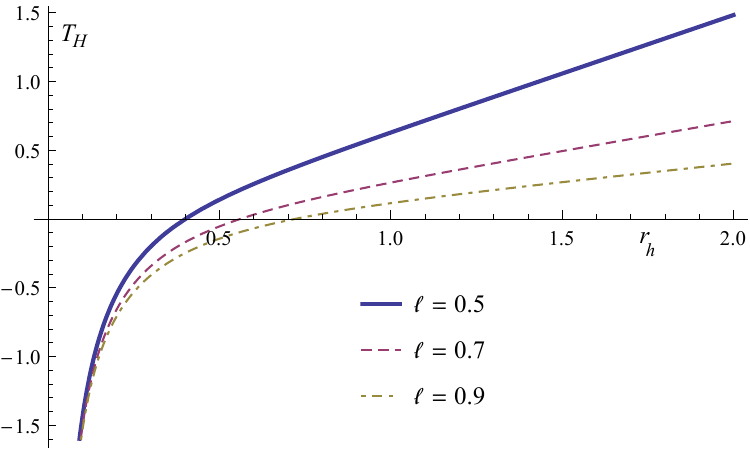}
 	\caption{Hawking temperature of the bilocal static black string as a function of the horizon event radius, for some values of $\ell$.}
 	\label{Hawking}
 \end{figure}

The Hawking temperature of the bilocal black string can be computed by means of $T_H=\frac{1}{4\pi}f'(r_h)$, where $r_h$ is the position of the external horizon. We cannot find analytically  $r_h$, but we can determine the value of $\mu$ as a function of such event horizon by making $f(r_h)=0$ and then substitute it in $f'(r)$.  In Fig. \ref{Hawking} we show the resulting temperature for some values of $\ell$. It is possible to see that for small $\ell$ we obtain $T_H\approx 2r_h/\ell^2$, as in the ordinary static black string case.

 Note that there is a critical event horizon position for which the Hawking temperature vanishes and the evaporation process halts giving rise to a remnant mass. For $\ell\ll r$, this critical horizon radius is given by
 \begin{equation}
 r^{rem}_h\approx\exp \Big\{-\frac{-3 \sqrt{2} \log \left[\left(3+2 \sqrt{2}\right) \ell^2\right]-8 \sqrt{2}+6+3 \sqrt{2} \log (8)}{6 \sqrt{2}}\Big\},
 \end{equation}
which corresponds to a remnant density of mass $\mu\approx 0.39$ in Planck units. Note that if $\ell\to 0$, $r_h^{rem}\to 0$, and there is no remnant in accordance with general relativity.

\subsection{Stability and energy conditions}
 Let us now investigate the stability of this black string.  We assume the existence of an anisotropic cosmological ideal fluid obeying a linear state equation $\tilde{p}(r)=\omega(r) \tilde{\rho}(r)$, where $\tilde{p}(r)$ is the energy pressure and $\omega(r)$ is a parameter so defined \cite{omega}.
  This fluid sustains the black string through the Einstein equation
 \begin{equation}
 G^{r}_{r}+\Lambda=\frac{\ell^2 f'(r)+2 r^2 f'(r)+2 r f(r)}{2 r \left(\ell^2+r^2\right)}+\Lambda=\kappa \omega(r) \tilde{\rho}(r),
 \end{equation}
 where $f(r)$ is given by Eq. (\ref{PotMetr}) and $\tilde\rho(r)$ by Eq. (\ref{Rho}).
 The numerical analysis of this equation shows that the fluid is only compatible with $\omega>-1$ as generally expected \cite{omega}.

 Now, we analyze the energy conditions for the fluid supporting the bilocal black string.
 Since the source is scattered in space as a result of nonlocality, we must find the regions where the Null Energy Conditions (NEC, $\rho+p_i\geq 0$), Weak Energy Conditions (WEC, $\rho\geq 0$, $\rho+p_i\geq 0$), Strong Energy Conditions (SEC, $\rho+p_i\geq 0$, $\rho+\sum p_i\geq 0$), and Dominant Energy Conditions (DEC, $\rho\geq 0$, $-\rho\geq p_i\geq\rho$) are satisfied. The index $i$ is for the spatial coordinates, $(1,2,3)\equiv (r,\phi,z)$.

\begin{figure}[h!]
\centering
       \includegraphics[width=0.6\textwidth]{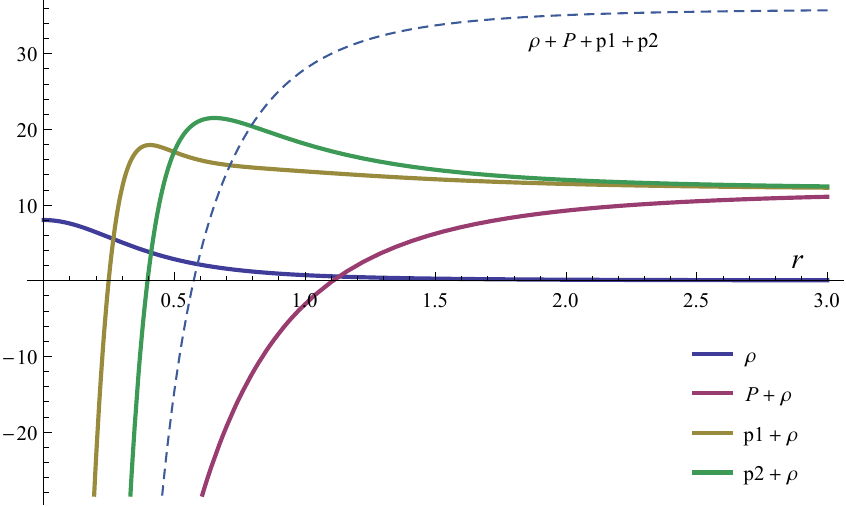}
           \caption{Energy density combined to the pressures (radial $p_r=P$ and lateral $p_{\phi}=p1$ and $p_z=p2$), as functions of the radial coordinate $r$, for $\ell=0.5$ and $\mu=2$, in Planck units.}
     \label{CondErg}
    \end{figure}

 The first thing to observe from Fig. \ref{CondErg} is that NEC, WEC, and SEC are respected for $r\gtrsim 1.16$.  Thus, exotic ($\omega<-1$) energy \cite{dark} nearby the object is not excluded by this analysis.

 \subsection{Stationary solution}

\begin{figure}[h!]
\centering
    \begin{minipage}{0.5\textwidth}  \nonumber
        \centering
        \includegraphics[width=0.9\textwidth]{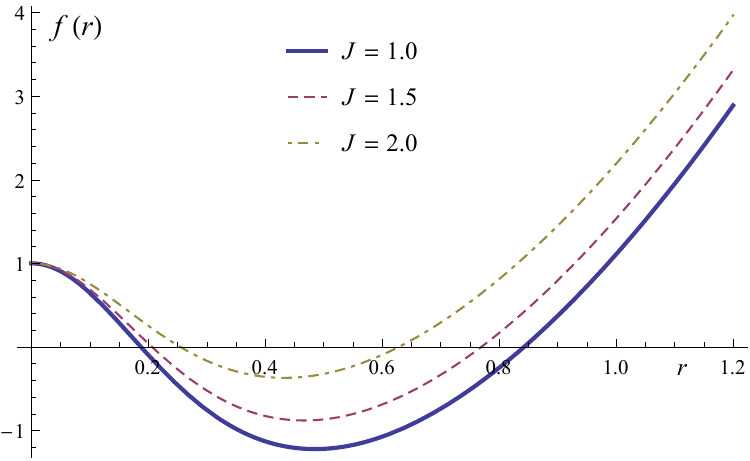}
    \end{minipage}\hfill
    \begin{minipage}{0.5\textwidth}  \nonumber
        \centering
        \includegraphics[width=0.9\textwidth]{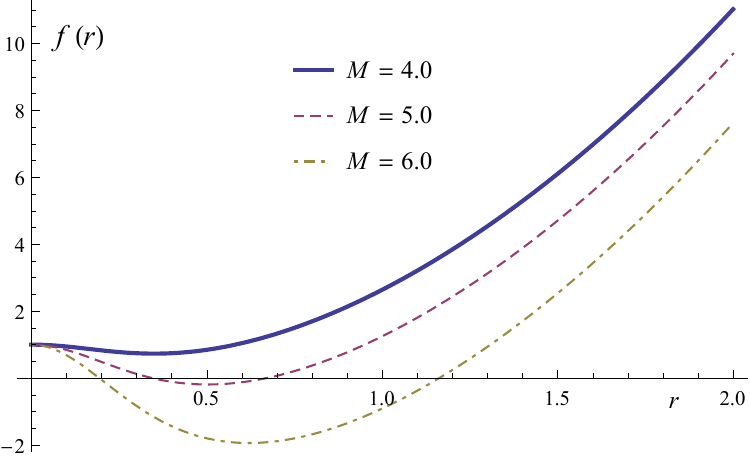}
    \end{minipage}\hfill
    \caption{Inverse radial coefficient of the stationary bilocal black string metric, as a function of the radial coordinate, for some values of angular momentum $J$ ($M=6.0$, left panel), and for some values of mass ($J=1.5$, right panel), with $\ell=0.5$, in Planck units.}
    \label{Stat}
\end{figure}

We now conceive the stationary counterpart of the black string through the transformation \cite{Lemos:1995cm}
\begin{eqnarray}
dt \to \lambda dt-\omega\ell^2 d\phi, \quad d\phi \to \lambda d\phi-\omega dt,\label{transf1}
\end{eqnarray}
where $\lambda$ and $\omega$ are constants to adjust. This yields
\begin{eqnarray}\label{stat1}
ds^2=-[f(r)\lambda^2-r^2\omega^2]dt^2-2[r^2-\ell^2f(r)]\lambda \omega d\phi dt \nonumber \\
-\frac{dr^2}{f(r)}+[r^2\lambda^2-f(r)\omega^2\ell^4]d\phi^2+\frac{r^2+\ell^2}{\ell^2}dz^2.
\end{eqnarray}
The identification with both black string mass $M$ and angular momentum $J$ can be made via the asymptotic approximation given by Eq. (\ref{AsymptoticMetric}). Following \cite{Lemos:1994xp} we obtain
\begin{eqnarray}
\lambda^2&=&\frac{M+ \sqrt{M^2-\frac{8 J^2}{9\ell^2}}}{-M+ 3\sqrt{M^2-\frac{8 J^2}{9\ell^2}}},\\
\omega^2\ell^2&=& \frac{2\left(M- \sqrt{M^2-\frac{8 J^2}{9\ell^2}}\right)}{-M+ 3\sqrt{M^2-\frac{8 J^2}{9\ell^2}}},
\end{eqnarray}
which implies
\begin{equation}
\kappa\mu=\frac{2\sqrt{2}}{4\ell\left[2-\sqrt{2}\log{(2\sqrt{2}\ell^2+3\ell^2)}\right]}\left(-M+ 3\sqrt{M^2-\frac{8 J^2}{9\ell^2}}\right).
\end{equation}
In Fig. \ref{Stat}, we depict the inverse radial coefficient $g_{rr}^{-1}$ of the stationary bilocal black string metric. It shows the behavior of the horizons with respect to the radial coordinate, for different values of the black string mass and angular momentum. Note that, for a given mass, the internal (external) horizon gets closer to (further from) the origin, the smaller (greater) the black string angular momentum. The opposite happens when we vary the mass for a fixed angular momentum. Notice also that for a sufficiently high (low) angular momentum (mass), the horizons disappear.

\section{Conclusions}

We have obtained a static uncharged black string solution to the modified Einstein equations of nonlocal gravity. Using an operator of order $1/2$ we have constructed an AdS-like solution which is regular at the origin (see Fig. 1).  This is confirmed by the finite value of the curvature scalar at this point, which is guaranteed provided a fundamental length is present. In the asymptotic limit, we  verified that the solution converges to the static black string of general relativity (see Eq. (\ref{AsymptoticMetric})).

We have shown that for a given value of $\Lambda$ the object allows up to two horizon radii whose positions depend on the mass density. We have found that the Hawking temperature presents the usual linear behaviour for large values of the event horizon. However, for a critical value of this horizon the Hawking temperature vanishes and there is a black string remnant (see Fig. \ref{Hawking}).

We have investigated the possibility of a perfect cosmological fluid to sustain the solution. In principle, this fluid is only compatible with $\omega>-1$ throughout space. However, by considering the energy conditions we found that the possibility of exotic energy is not excluded  nearby the black string.

Finally, following \cite{Lemos:1994xp} we found the stationary black string solution. As in the static case, the rotating string presents up to two horizons. Their position as a function of the object's mass and angular momentum, as well as the regularity of the stationary solution, is shown in Figs. \ref{Stat}.

\section*{Acknowledgments} The authors would like to thank Conselho Nacional de Desenvolvimento Cient\'{i}fico e Tecnol\'{o}gico (CNPq) for the partial financial support. MSC is under the CNPq research project No. 315926/2021-0.

\section*{Data availability statement}

Data sharing is not applicable to this article as no datasets were generated or analysed
during the current study.



\end{document}